\documentclass[ALICE,manyauthors]{cernphprep}

\usepackage{graphicx}
\usepackage{hyperref}   %% produces clickable links in xdvi
\usepackage{flafter}
\usepackage{amsmath}
\usepackage{float}
\usepackage{wrapfig}
\usepackage{subfigure}
\usepackage{epsfig}
\usepackage{epstopdf}
\usepackage{fancyhdr}
\usepackage{mathtools}
\usepackage{amssymb}
\usepackage{multicol}
\usepackage{multirow}
\usepackage{color}
\usepackage[modulo]{lineno}
\usepackage{array}
\usepackage{soul}
\usepackage{caption}
\usepackage{enumitem}
\usepackage[titletoc]{appendix}
\usepackage{booktabs}
\usepackage[usenames,dvipsnames]{xcolor}
\usepackage{cite}
\usepackage{bm}
\usepackage{svn-multi}
\svnid{$Id: dijetKt_paper.tex 504 2015-03-03 08:01:03Z mverweij $}

\DeclareBoldMathCommand\boldlangle{\left\langle}
\DeclareBoldMathCommand\boldrangle{\right\rangle}

\newcommand{\rom}[1]{{\mathrm{#1}}}   % switch to rom in math 

\newcommand{\kt}{{$k_{\rm T}$}}

\newcommand{\kty}{\ensuremath{k_{\rm Ty}}}
\newcommand{\ktyabs}{\ensuremath{|k_{\rm Ty}|}}

\newcommand{\PbPb}{Pb--Pb}
\newcommand{\pp}{pp}
\newcommand{\pPb}{p--Pb}
\newcommand{\sNN}{\ensuremath{\sqrt{s_\mathrm{NN}}}}
\newcommand{\GeV}{GeV}
\newcommand{\GeVc}{\text{\GeV/}\ensuremath{c}}

\newcommand{\deltaphi}{\ensuremath{\Delta\varphi_{\rom{dijet}}}}

\newcommand{\pt}{\ensuremath{p_\rom{T}}}

\newcommand{\deltapt}{\ensuremath{\rom{\delta}\pt}}

\newcommand{\ptjet}{\ensuremath{p_\rom{T,jet}}}

\newcommand{\ptjetfull}{\ensuremath{p_\rom{T,jet}^\rom{ch+ne}}}

\newcommand{\ptjetchassoc}{\ensuremath{p_\rom{T,assoc\;jet}^\rom{ch}}}

\newcommand{\pttrack}{\ensuremath{p_\rom{T,track}}}

\newcommand{\chisq}{\ensuremath{\chi^{2}}}

\newcommand{\dphidijet}{\ensuremath{\Delta\varphi_{\rom{dijet}}}}

\newlength{\myfigwidth}
\begin{document}

%%%%%%%%%%%%%%%  Title page %%%%%%%%%%%%%%%%%%%%%%%%
%
\begin{titlepage}
\PHyear{2015}           % % required, will be obtained from PH
\PHnumber{041}             % % required, will be obtained from PH
\PHdate{24 Feb}  
%

%%% Put your own title + short title here:
\title{Measurement of dijet $\mathbf{{\textit{k}}_{T}}$ in \pPb{} collisions at $\mathbf{\sqrt{{\textit{s}}_{NN}}=5.02}$ TeV}% \newline \newline \version}
\ShortTitle{Measurement of dijet \kt{} in \pPb{} collisions at $\sNN=5.02$ TeV}   % appears on right page headers
%
%%% Do not change the next lines!
\Collaboration{ALICE Collaboration%
         \thanks{See Appendix~\ref{app:collab} for the list of collaboration
                      members}}
\ShortAuthor{ALICE Collaboration}      % appears on left page headers, do not change
%
%\linenumbers

\begin{abstract}
A measurement of dijet correlations in \pPb{} collisions at $\sNN=5.02$ TeV with the ALICE detector is presented. Jets are reconstructed from charged particles measured in the central tracking detectors and neutral energy deposited in the electromagnetic calorimeter. 
The transverse momentum of the full jet (clustered from charged and neutral constituents) and charged jet (clustered from charged particles only) is corrected event-by-event for the contribution of the underlying event, while corrections for underlying event fluctuations and finite detector resolution are applied on an inclusive basis. 
A projection of the dijet transverse momentum, $\kty = \ptjetfull \; \rom{sin}(\Delta\varphi_{\rom{dijet}})$ with $\Delta\varphi_{\rom{dijet}}$ the azimuthal angle between a full and charged jet and $\ptjetfull$ the transverse momentum of the full jet, is used to study nuclear matter effects in \pPb{} collisions. This observable is sensitive to the acoplanarity of dijet production and its potential \mbox{modification} in \pPb{} collisions with respect to pp collisions. Measurements of the dijet \kty{} as a function of the transverse momentum of the full and recoil charged jet, and the event multiplicity are presented. No significant modification of \kty{} due to nuclear matter effects in \pPb{} collisions with respect to the event multiplicity or a PYTHIA8 reference is observed.
\end{abstract}
\end{titlepage}
\setcounter{page}{2}

%\linenumbers
\section{Introduction}
Dijets produced in $2\rightarrow2$ leading-order (LO) scattering processes are balanced in transverse momentum and back-to-back in azimuth. In proton-proton collisions a small acoplanarity appears due to intrisic transverse momentum \kt{} from partonic Fermi motion \cite{Feynman:1977yr} and initial state gluon radiation \cite{PhysRevLett.81.2642,PhysRevD.59.074007}. At large momentum transfer between the incoming partons, the phase space for hard gluon radiation in the parton shower or from next-to-leading-order (NLO) processes increases, resulting in acoplanarity of the dijet system \cite{Abazov:2004hm,Khachatryan:2011zj,daCosta:2011ni}. This also results in an imbalance of the jet transverse momenta also referred to as a broadening of the dijet transverse momentum. The relative contribution of hard QCD radiation to the dijet \kt{} can be varied by applying kinematic and acceptance selections to the dijet sample.

In \pPb{} collisions the dijet kinematics are potentially modified due to nuclear matter effects which are expected to induce a momentum imbalance and acoplanarity of dijet pairs with respect to pp collisions, so-called transverse momentum broadening \cite{Albacete:2013ei}. For instance, multiple scatterings inside the nucleus of the initial- and final-state partons in hard scatterings can lead to such a transverse momentum broadening.

In heavy-ion collisions, jets produced in hard scattering processes are used to probe the properties of the produced medium. Highly energetic partons propagate through the medium, which modifies the parton shower resulting in a modified fragmentation pattern of the final hadronization products \cite{Gyulassy:1990ye,Baier:1994bd}. Heavy-ion jet measurements are compared to measurements in pp collisions to determine the effect of hot nuclear matter on jet observables \cite{Aad:2014bxa,Aad:2014wha,Chatrchyan:2014ava,Chatrchyan:2013kwa}. In the context of such studies, measurements in \pPb{} collisions serve as a benchmark to study hard scattering processes in a nuclear target.

Measurements presented in \cite{Chatrchyan:2014hqa} of the dijet transverse momentum imbalance and dijet azimuthal angle distributions show results which are comparable to results obtained with \pp{} data and independent of the event activity. This letter presents a measurement of dijet acoplanarity in \pPb{} collisions at $\sNN=5.02$ TeV, recorded with the ALICE detector at the Large Hadron Collider (LHC). The jet azimuthal correlations are measured at mid-rapidity for jet transverse momentum between 15 and 120 \GeVc. Jets entering in the acceptance of electomagnetic calorimeter are reconstructed from charged and neutral particles (full jet) while the recoil jet is reconstructed from charged particles only (charged jet). Measurements are presented as a function of the full and associated charged jet transverse momentum in two event \mbox{multiplicity} classes which are correlated to the centrality of the \pPb{} collisions \cite{Adam:2014qja}.

\section{Experimental Setup and Data Sample}\label{sec:DataSample}
Collisions of proton and lead beams were provided by the LHC in the first months of 2013. The beam energies were 4 TeV for the proton beam and 1.58 TeV per nucleon for the lead beam, resulting in collisions at a center of mass energy $\sNN = 5.02$ TeV. The nucleon-nucleon center-of-mass system moves in rapidity with respect to the ALICE reference frame by -0.465 in the direction of the proton beam \cite{ALICE:2012xs}. In the following $\eta$ refers to the pseudorapidity in the ALICE reference frame. %defined such that the proton moves in positive $\eta_{\rom{cms}}$.

The V0 detectors, two arrays of scintillator tiles covering the full azimuth within $2.8<\eta<5.1$ (V0A) and $-3.7<\eta<-1.7$ (V0C), were used for online minimum bias event triggering, offline event selection and characterization of events in different particle multiplicity classes. The minimum bias trigger required a signal from a charged particle in both the V0A and V0C. The total integrated luminosity of the minimum bias event sample is 37 $\rom{\mu b}^{-1}$. 

The electromagnetic calorimeter (EMCal) in ALICE \cite{oai:arXiv.org:1008.0413} covers 100 degrees in azimuth, $1.4<\varphi<\pi$, and $|\eta|<0.7$. 
For the analyzed data set an online jet patch trigger of $32\times32$ adjacent towers, corresponding to an area of approximately $0.2$ $\mathrm{rad}$ was used. This jet patch trigger fired if an integrated patch energy of at least $10$~\GeV{} (low-energy trigger) or $20$~\GeV{} (high-energy trigger) was found. The low-energy triggered event sample provided a significant overlap in jet energy between the minimum bias and high-energy trigger event samples, allowing assessment of the trigger biases. The event sample obtained with the low-energy trigger corresponds to a total integrated luminosity of 21 $\rom{\mu b}^{-1}$. The event sample with the high energy threshold has a total integrated luminosity of 1.6 $\rom{nb}^{-1}$.

The position of the primary vertex was determined using reconstructed charged particle tracks in the ALICE tracking systems, Inner Tracking System (ITS) \cite{Aamodt:2010aa} and Time Projection Chamber (TPC)  \cite{Alme:2010ke}. The algorithm to reconstruct the primary vertex is fully efficient for events with at least one primary track within $|\eta|<1.4$ \cite{Abelev:2014ffa}. To ensure a high tracking efficiency uniform in $\eta$, events are accepted if the coordinate of the vertex along the beam direction is within $\pm10$ cm from the center of the detector.

The total event sample is divided into two multiplicity classes based on the total charge deposited in the V0A detector \cite{Abelev:2014mva}. For the data sample used in this analysis, the V0A detector is located in the direction of the Pb remnants and thus sensitive to the fragmentation of the nucleus limiting a correlation in the definition of the multiplicity class with the dijet measurement at midrapidity. Two multiplicity classes 0-40\% and 40-100\% are used in this analysis. The higher multiplicity class \mbox{(0-40\%)} corresponds to $\langle\rom{d}N/\rom{d}\eta\rangle_{|\eta|<0.5} = 37.2 \pm 0.8$ and the lower multiplicity class \mbox{(40-100\%)} to $\langle\rom{d}N/\rom{d}\eta\rangle_{|\eta|<0.5} = 9.4 \pm 0.2$.

\section{Jet reconstruction and dijet $\mathbf{{\it k}_{Ty}}$}

\subsection{Jet Reconstuction}
Jets are reconstructed with the anti-\kt{} jet algorithm of the FastJet package \cite{Cacciari2011, Cacciari2006} combining charged tracks measured in the central tracking detectors, ITS and TPC, and neutral fragments measured with the EMCal \cite{oai:arXiv.org:1008.0413}. Tracks from the combined ITS and TPC track reconstruction algorithm are used. Quality criteria for track selection follow the same strategy as in \cite{Abelev:2012ej}. The tracking efficiency is 70\% for tracks with a transverse momentum $\pttrack=0.15$~\GeVc{} and increases to 85\% at $\pttrack=1$~\GeVc{} and above. The \pt{} resolution of tracks is 0.8\% (3.8\%) for $\pttrack=1$~\GeVc{} ($50$~\GeVc). EMCal clusters are formed by a clustering algorithm that combines signals from adjacent EMCal towers, with cluster size limited by the requirement that each cluster contains only one local energy maximum. Energy deposited by charged particles in the EMCal is subtracted from the measured energy in the EMCal clusters which prevents counting 
the charged energy twice \cite{Abelev:2013fn,Abelev:2014ffa}. ALICE also reconstructs jets from charged particles only. These jets are referred to as `charged jets', while jets reconstructed from charged and neutral fragments are called `full jets' in this letter.

In this analysis, anti-\kt{} jets are reconstructed using the boost-invariant \pt{} recombination scheme and a jet resolution parameter of $R=0.4$. A jet is only accepted if it is fully contained in the acceptance in which the constituents are measured: for charged jets in the full azimuth and $|\eta_\rom{jet}^\rom{ch}|<0.9-R$ while for full jets $1.4+R<\varphi_\rom{jet}^\rom{ch+ne}<\pi-R$ and $|\eta_\rom{jet}^\rom{ch+ne}|<0.7-R$. It was verified that reducing the acceptance with 0.05 on all edges, in $\eta_\rom{jet}$ and $\varphi_\rom{jet}$, has a negligible effect on the measurement. Tracks with $\pttrack>0.15$~\GeVc{} and neutral constituents with $E_{\rm T}>0.3$ GeV are considered. The minimum required area for jets with a resolution parameter $R=0.4$ is equal to $0.3$ ($\approx60\%$ of the area of a rigid cone with $R=0.4$). This selection does not affect the jet finding efficiency for jets (full and charged) with transverse momentum $p_\rom{T,jet} > 15$~\GeVc. In addition, jets containing a track with $\pttrack>100$~\GeVc{}, for which the track momentum resolution exceeds 6.5\%, are tagged and rejected. This last requirement has negligible effect in the reported range of jet momenta. The measurement is corrected to particle level as will be explained in Section \ref{sec:corrections}.

The measured transverse momentum of the anti-\kt{} jet is corrected for the contribution of the underlying event by subtracting the average background momentum density, $\rho$ for full jets and $\rho_{\rom{ch}}$ for charged jets, multiplied by the area of the considered jet. 
The contribution of the underlying event to the charged jets is estimated using clusters reconstructed with the \kt{} jet algorithm using only charged tracks. This is achieved by calculating event-by-event the median charged background density, $\rho_{\rom{ch}}$, from all \kt{} clusters in the event with in addition a correction for the sparsely populated \pPb{} events \cite{Chatrchyan:2012tt,Adam:2015hoa}. 
The average $\rho_{\rom{ch}}$ in minimum-bias events is equal to $1.9$~\GeVc{} for the 0--40\% multiplicity class and $0.7$~\GeVc{} for the 40--100\% event multiplicity class with a  negligible statistical uncertainty in both cases. Finally, $\rho_{\rom{ch}}$ is multiplied by a scale factor to account for the neutral energy to estimate $\rho$. The scale factor is determined by measuring the ratio between the energy of all the EMCal clusters and the charged tracks pointing into the EMCal acceptance in the minimum-bias event sample. The extracted scale factor $1.28$ is independent of event multiplicity. The influence of background fluctuations is quantified and corrected for on an inclusive basis, see Section \ref{sec:corrections}.

\subsection{Dijet $\mathbf{{\it k}_{Ty}}$}
Each measured full jet is correlated with the charged jet of highest transverse momentum in the opposite hemisphere. Only pairs for which the full jet has a larger transverse momentum than the associated charged jet are considered. Furthermore only dijets pairs with $|\Delta \varphi_{\rom{dijet}}-\pi| < \pi/3$, with $\Delta\varphi_{\rom{dijet}}$ the angle between the jet axis of the full and charged jet, are considered in the analysis. The selection in $\Delta \varphi_{\rom{dijet}}$ rejects 5--8\% of the dijet pairs depending on the kinematic selection of the full and associated charged jet. 
The azimuthal acoplanarity of dijets is studied by measuring the transverse component of the \kt{} vector of the dijet system, \kty, defined as
\begin{equation}
 \kty = \ptjetfull \; \rom{sin}(\Delta\varphi_{\rom{dijet}}),
\end{equation}
with \ptjetfull{} the transverse momentum of the full jet. It should be noted that this definition differs from the one used in previous publications, for example \cite{Angelis1980163}. 
Since $\rom{d}N\rom{/}\rom{d}\kty$ is a symmetric distribution around zero, \ktyabs{} is reported throughout the paper. 
For events from the minimum-bias sample full jets with $\ptjetfull>20$~\GeVc{} are considered, while in the jet-triggered data samples (see Section \ref{sec:DataSample}) only jets with $\ptjetfull>40$~\GeVc{} for the low energy trigger and $\ptjetfull>60$~\GeVc{} for the higher energy trigger are used. In these kinematic regimes the triggers are fully efficient and no fragmentation bias is observed with respect to the minimum-bias jet sample. The \ktyabs{} distributions are reported at particle level involving a correction for detector effects and \pt-smearing due to the underlying event, see Section \ref{sec:corrections}.

The dijet sample is biased due to the requirement that the full jet has a larger transverse momentum than the associated charged jet, while the full jet momentum is used to estimate \kty. 
In an unbiased measurement, the full jet would correspond to the leading jet of the event in only 50\% of the cases. 
A PYTHIA \cite{Sjostrand2006,Sjostrand:2007gs} study was performed in which the particle-level jet was defined as the jet containing all final state particles (no kinematic selection on constituents and full azimuthal acceptance). Applying the selection of this analysis to detector-level reconstructed jets, results in a correct tagging of the leading jet in 70\% of the dijet events. This results in a slightly harder \ktyabs{} distribution with a 10\% smaller yield at low \ktyabs{} and 20\% higher yield at large \ktyabs. The results of the \mbox{\pPb{}} data analysis will be compared to particle-level PYTHIA with the same dijet selection incorporating the mentioned bias.

The dijet acoplanarity is measured as a function of the transverse momentum of the full jet while the kinematic interval of the associated charged jet is also varied to explore \kty{} for more or less balanced dijets in transverse momentum. In addition \kty{} distributions are also presented for two event multiplicity classes.

\subsection{Corrections and Systematic Uncertainties}\label{sec:corrections}
The measured dijet \ktyabs{} distributions are corrected to the particle level, defined as the dijet \ktyabs{} from jets clustered from all prompt particles produced in the collision including all decay products, except those from weak decays of light flavor hadrons and muons. Both full and charged jets are accepted at particle level in the full azimuthal acceptance and in the pseudorapidity range of $|\eta_{\rom{jet}}|<0.5$. The correction to particle level is based on a data-driven method to correct for the influence of the underlying event fluctuations and on simulated PYTHIA events (tune Perugia-2011 \cite{Skands:2010ak}) transported through the ALICE detectors layout with GEANT3 \cite{Brun1994}. The correction procedure takes into account the \pt{} and angular resolution of the measured dijets.

Detector-level jets are defined as jets reconstructed from reconstructed tracks and EMCal clusters after subtraction of the charged energy deposits. The jet energy scale and resolution are affected by unmeasured particles (predominantly $\rom{K}^{0}_{\rom{L}}$ and neutrons), fluctuations of the energy deposit by charged tracks in the EMCal, the EMCal energy scale and the charged particle tracking efficiency and \pt{} resolution. 
A response matrix as a function of \ptjet{} of the full and associated charged jet, \dphidijet{} and \kty{} is created after matching the detector-level to the particle-level jets as described in \cite{Abelev:2013kqa}.

The \pt-smearing due to fluctuations of the underlying event is estimated with the random-cones technique which is also applied in the analysis of \PbPb{} data \cite{Abelev:2012ej}. Cones with a radius equal to the resolution parameter $R$ are placed in the measured \pPb{} events at random positions in the $\eta-\varphi$ plane ensuring the cone is fully contained in the detector acceptance. The fluctuations of the background are characterized by the difference between the summed \pt{} of all the tracks and clusters in the random cone (RC) and the estimated background:
$\deltapt = \sum_{i}^{\rom{RC}}p_{\rm T,i} - A \cdot \rho$, where $A$ is the area of the random cone ($A = \pi R^2$) and the subscript $i$ indicates a cluster or track pointing inside the random cone. A random cone can overlap with a jet but to avoid oversampling in small systems like \pPb{}, a partial exclusion of overlap with the leading jet in the event is applied. This is achieved by excluding random cones overlapping with a leading jet with a given probability, $p=1/N_{\rom{coll}}$ where $N_{\rom{coll}}$ is the number of binary collisions. $N_{\rom{coll}}$ is taken from estimates applying a Glauber fit to the multiplicity measured in the V0A detector resulting in values between $14.7$ and $1.52$ depending on the event activity measured in the V0A detector. The width of the background fluctuations for full (charged) jets varies between $2.12$ ($1.59$) and $0.73$ ($0.56$) \GeVc{} depending on the multiplicity of the event.

The influence of background fluctuations is added to the response extracted from detector simulation through a Monte Carlo model assuming that the background fluctuation for the full and associated charged jet are uncorrelated within 20\% wide bins of V0A multiplicity classes. Within these selected multiplicity classes the variation of the background fluctuations is negligible. Since the Monte Carlo model does not generate full events and only accounts for the \deltapt{} smearing on a jet-by-jet basis, additional jet finding inefficiencies and worsening of angular resolution due to the background fluctuation are not taken into account. These effects are negligible since the contribution of the underlying event to the jets in \pPb{} collisions is small. 
No correction for the angular resolution of the charged jet due to missing neutral fragments is applied. This effect increases the width of \deltaphi{} by $\sim0.03$ and is present, and of the same magnitude, in the \pPb{} data and the PYTHIA reference.

The most probable correction to the jet energy, taking into account detector effects and background fluctuations, for fully reconstructed jets is 28\% at $\ptjetfull=20$ \GeVc{} and decreases to 20\% for jets with $\ptjetfull>40$ \GeVc. The uncertainty on the jet energy scale is evaluated by changing the tracking efficiency in data and full detector simulation \cite{ALICE:2014dla}, varying the double counting correction for the hadronic energy deposit in the EMCal and by using different estimates of the underlying-event fluctuations. The final uncertainty on the jet energy scale is 4\%. The jet energy resolution for full jets is 22\% at $\ptjetfull=20$ \GeVc{} and decreases gradually to 18\% at $\ptjetfull=120$ \GeVc. The influence of the uncertainties on the jet energy scale and resolution on the dijet \ktyabs{} measurement are discussed in the following.

The measured \ktyabs{} distributions are corrected to the particle level by applying bin-by-bin correction factors, which are parametrized by a linear fit to the ratio between the particle- and detector-level \ktyabs{} distributions for a given dijet selection. The correction factors take into account the effects of feed-in and feed-out of the selected kinematic and angular intervals of the full and associated charged jets. These effects slightly change the shape of the \ktyabs{} distributions resulting in correction factors which vary between $0.9$ for small \ktyabs{} to $1.2$ at large \ktyabs. The correction is relatively small, because while feed-in from lower \ptjetfull{} narrows the \ktyabs{} distribution, feed-in from higher \ptjetfull{} broadens the distribution resulting in a cancellation. Similarly the feed-out to high and low \ptjetfull{} has a small effect on the observable. By using a linear fit to the correction factors the statistical fluctuations of the detector simulation are not propagated 
to the 
measurement. The 95\% confidence limit of the parametrization using the linear fit is included in the systematic uncertainty of the measurement. Correction factors are extracted as a function of V0A event multiplicity class and kinematic intervals of the full and associated charged jet.

The dominant systematic uncertainty on the measurement originates from the extraction of bin-by-bin correction factors. The uncertainty of the parametrization of the correction factors results in 10--20\% correlated systematic uncertainty on the dijet \ktyabs{} yields. An additional 2.5\% uncertainty arises from the uncertainty on the tracking efficiency which is 4\%. Systematic uncertainties originating from the charged hadron energy deposit in EMCal towers, background fluctuations and the average momentum density $\rho$ were evaluated and found to be negligible.
\section{Results}

The dijet \ktyabs{} distributions are presented as functions of the full-jet transverse momentum, the transverse momentum of the associated charged jet and event multiplicity classes. The results are compared to the predictions of the PYTHIA8.176 event generator with tune 4C and $K=0.7$ \cite{Sjostrand2006,Sjostrand:2007gs}. This tune has been found to give a reasonable description of jet production at the LHC. The final state particles are shifted in pseudorapidity with $\eta_{\rm shift}=-0.465$ to mimic the rapidity shift of the laboratory frame due to the energy difference of the proton and Pb beams.

\subsection{Evolution with full-jet transverse momentum}\label{sec:EvoFullJetPt}
\begin{figure}[!ht]
\includegraphics[width=\linewidth]{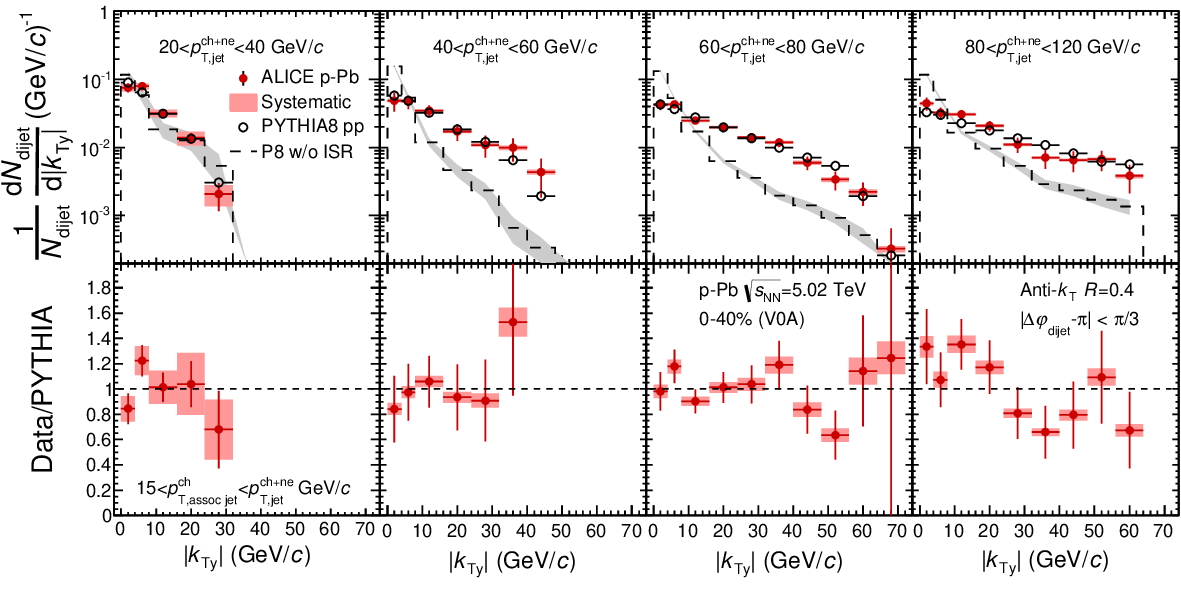}
\caption{\label{fig:KtPtTrigPythiaFuCh}Dijet \ktyabs{} distributions in \pPb{} collisions in the 0-40\% V0A multiplicity event class for several kinematic intervals of the full jet (\ptjetfull). The measurement is compared to PYTHIA8 (tune 4C, $K=0.7$) with and without initial state radiation. The lower panels show the ratio between the measurement and PYTHIA8 including initial state radiation.
}
\end{figure}

Figure \ref{fig:KtPtTrigPythiaFuCh} shows the corrected \ktyabs{} distributions for several kinematic intervals for the full jet, from $\ptjetfull=20$~\GeVc{} to $\ptjetfull=120$~\GeVc, in the 0--40\% V0A multiplicity class. The associated charged jet has a minimum transverse momentum, \ptjetchassoc, of 15~\GeVc{} and is always of lower transverse momentum than the full jet. The mean \ktyabs{} increases with the transverse momentum of the full jet.
Increasing the transverse momentum of the full jet extends the kinematic reach of \ktyabs{} by opening phase-space for more gluon radiation. This results in a harder \ktyabs{} distribution which drops at large \ktyabs{} because the kinematic limit $\ktyabs_{\rm max} = p_{\rom{T,jet,max}}^{\rom{ch+ne}}\;\rom{sin}(2\pi/3)$ is reached. 
The \pPb{} data points and the PYTHIA8 calculation show a similar dependence on \ptjetfull. 
The lower panels of Fig.~\ref{fig:KtPtTrigPythiaFuCh} show the ratio between data and PYTHIA8, including initial state radiation (ISR), which is observed to be consistent with unity for all transverse momentum ranges studied. In the upper panels PYTHIA without the initial state radiation option is shown in addition (dashed line). Without ISR the amount of QCD radiation (which includes NLO corrections) is reduced, resulting in a steeper \ktyabs{} spectrum. The effect is most pronounced for the $\ptjetfull>40$~\GeVc{} where the \pPb{} measurement is in agreement with full PYTHIA simulation but differs significantly from PYTHIA without ISR. This observation suggests that the dijet \ktyabs{} spectrum for large $Q^2$ processes is highly sensitive to the increased available phase-space of QCD radiation processes. 
Measurements presented in \cite{Chatrchyan:2014hqa} of the dijet transverse momentum imbalance for more energetic jets than the measurement presented here also show results which are comparable to simulated \pp{} reference and independent of the forward transverse energy.

\subsection{Evolution with event multiplicity and $\mathbf{{\it p}_{T,assoc\;jet}^{ch}}$}
\begin{figure}[!ht]
 \centering
 \includegraphics[width=\linewidth]{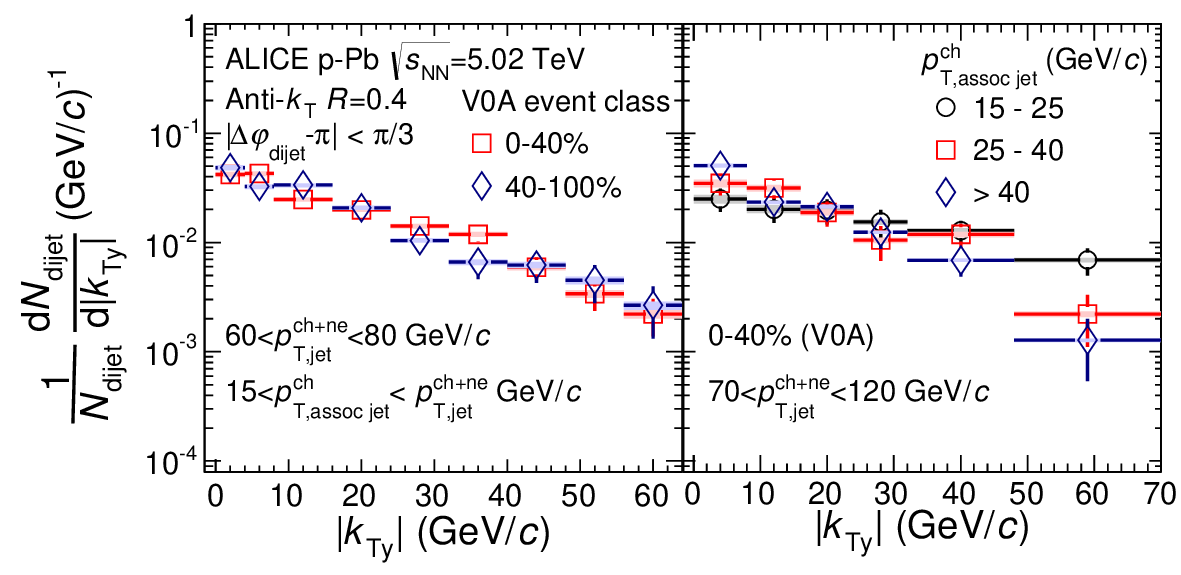}
 \caption{\label{fig:KtMultPtAss}Distributions of \ktyabs{} for two V0A event classes (left panel) and three \ptjetchassoc{} ranges (right panel).}
\end{figure}

In addition to the measurement of \ktyabs{} in the highest multiplicity \pPb{} events, the \ktyabs{} distribution is also measured in the lower multiplicity V0A event class \mbox{40-100\%}. If strong nuclear effects are present they are expected to be stronger in the high multiplicity events due to the larger number of participants in the collision. A comparison is shown in the left panel of Fig.~\ref{fig:KtMultPtAss}. The systematic uncertainties between the two measurements are fully correlated since they originate from the uncertainty on the jet energy scale of the full jet. The consistency between the \ktyabs{} distributions in the high and low multiplicity event class was evaluated by taking the ratio and performing a constant fit taking into account only the statistical errors. The fit is within 1.2$\sigma$ consistent with unity. 
%The \ktyabs{} distributions in the high and low multiplicity event class are within 1.2$\sigma$ consistent with each other. 
This result shows that in the measured kinematical region, possible nuclear matter effects and/or shadowing in the \ktyabs{} distributions in \pPb{} collisions are not observed for dijets at midrapidity.

The sensitivity to dijet acoplanarity is enhanced by selecting more \pt{} imbalanced jet pairs. 
The \ktyabs{} distribution for full jets with $70<\ptjetfull<120$~\GeVc{} for various \ptjetchassoc{} ranges is shown in the right panel of Fig.~\ref{fig:KtMultPtAss}. 
The \ktyabs{} distribution tends to become steeper if jets are more balanced indicating that the influence of QCD radiation decreases. This behavior supports the previous observation (Sec.~\ref{sec:EvoFullJetPt}) that the dijet \ktyabs{} observable for highly energetic jets is over a wide range of \ktyabs{} mainly sensitive to QCD radiation processes rather than elastic scatterings.

\subsection{Evolution and characterization via $\mathbf{\langle |{\it k}_{Ty}| \rangle}$}
The measured \ktyabs{} distributions are further characterized by reporting the mean ($\langle\ktyabs\rangle$) of the distribution. To avoid that the extracted moment is biased by statistical fluctuations for large values of \ktyabs, the distributions are extrapolated using a template generated with PYTHIA8 (tune 4C, $K=0.7$), which agrees well with the p--Pb measurement (see Fig.~\ref{fig:KtPtTrigPythiaFuCh}). The PYTHIA \ktyabs{} distribution is normalized to minimize the \chisq{} between data and PYTHIA. The transition from the data to the normalized template is fixed at 60\% of the kinematic limit $\ktyabs_{\rm max}$. The transition point is varied to estimate the systematic uncertainty from this extrapolation procedure. In addition, the normalization of the PYTHIA template is varied by one standard deviation of the fit uncertainty. This results in an additional systematic uncertainty on the extraction of $\langle\ktyabs\rangle$. For low \ptjetfull{} the uncertainty on the extracted mean is 
equal to 2.9\% and increases to 8.1\% for the highest \ptjetfull{} values.

\begin{figure}[!ht]
\includegraphics[width=0.48\linewidth]{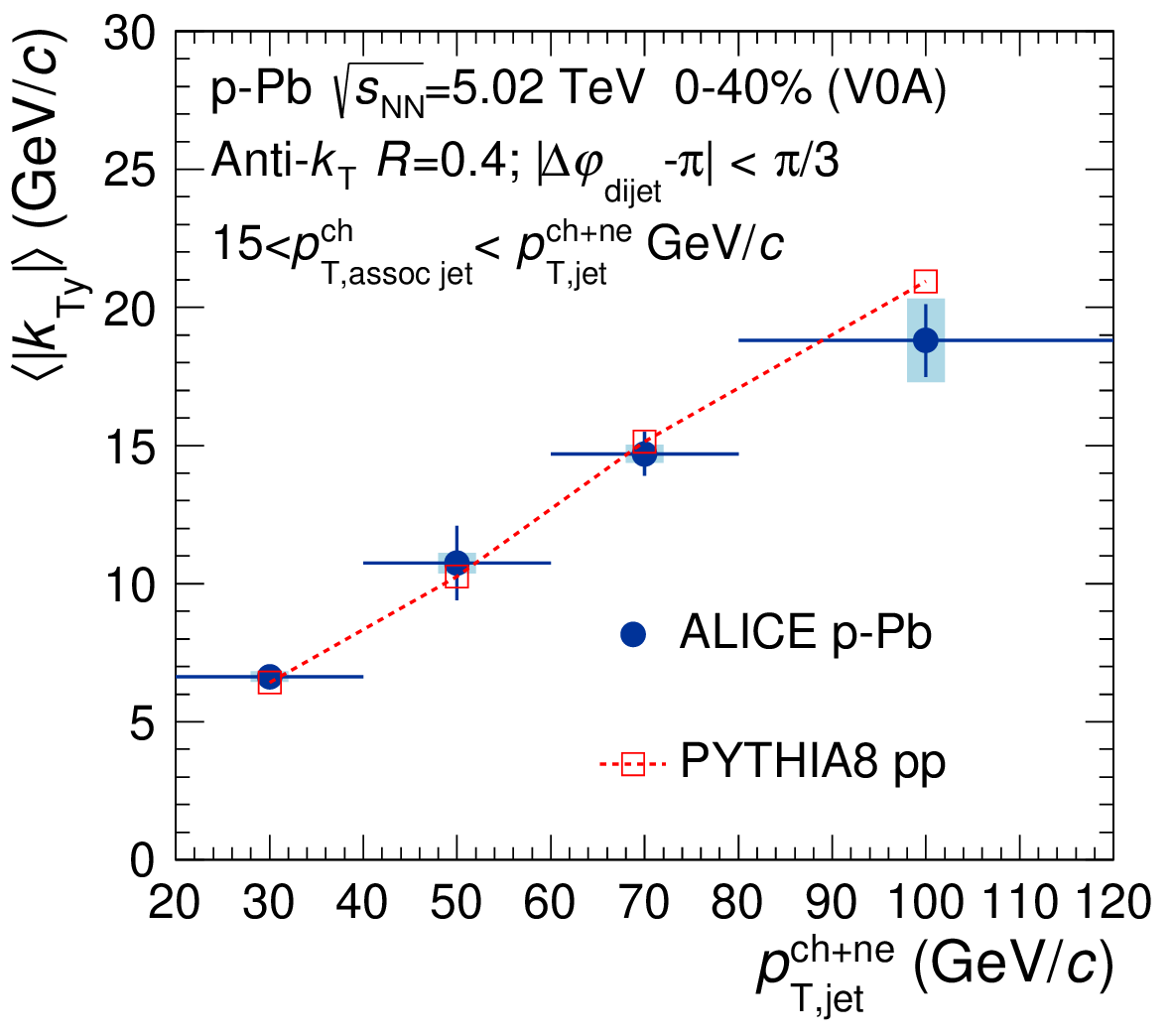}
\includegraphics[width=0.48\linewidth]{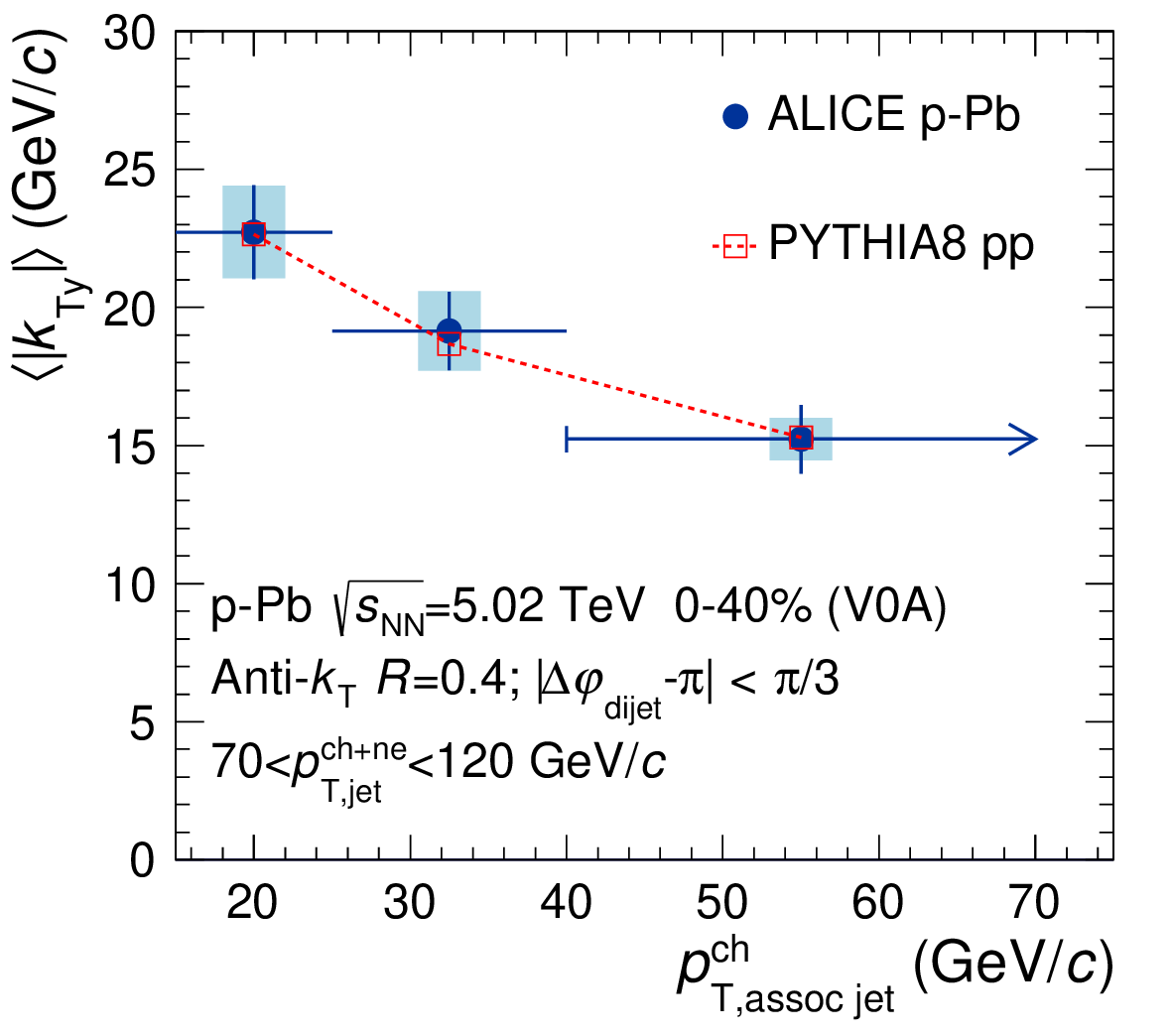}
\caption{\label{fig:MeanKtPythiaFuCh}Mean of the \ktyabs{} distributions as a function of the full jet transverse momentum \ptjetfull{} (left) and the associated charged jet transverse momentum \ptjetchassoc{} (right) compared to PYTHIA8.
}
\end{figure}
The left panel of Fig.~\ref{fig:MeanKtPythiaFuCh} shows the mean of the measured \ktyabs{} distributions as a function of the full jet transverse momentum and is compared to the PYTHIA values.
The measured moment in \pPb{} collisions agrees within the uncertainties of the measurement with the PYTHIA8 expectation. The mean increases with \ptjetfull{} since the additional \kt{} due to radiative QCD processes increases with \ptjetfull. 

\begin{table}[!ht]
 \begin{center}
 \renewcommand{\arraystretch}{1.3}
  \begin{tabular}{>{\small}l|>{\small}c>{\small}c>{\small}c}
 & 0-40\% & 40-100\% & PYTHIA8 pp \\
 \hline
 $\langle \ktyabs \rangle$ (\GeVc) & $14.7 \pm 0.8 \pm 0.3$ & $13.6 \pm 1.1 \pm 0.5$ & $15.1 \pm 0.1$ \\
  \end{tabular}
 \end{center}
\caption{\label{tab:MomentsV0AEventClass}Mean of the \ktyabs{} distributions for $60<\ptjetfull<80$~\GeVc{} and $15<\ptjetchassoc<\ptjetfull$~\GeVc{} in a high (0-40\%) and low (40-100\%) V0A multiplicity event class. The first quoted uncertainty is statistical while the second is systematic. The last column corresponds to the values from the PYTHIA8 calculation at particle level with the same kinematic selection. The uncertainty on the PYTHIA calculation is statistical.}
 \end{table}
The right panel of Fig.~\ref{fig:MeanKtPythiaFuCh} shows the evolution of $\langle\ktyabs\rangle$ as a function of \ptjetchassoc{} for $70<\ptjetfull<120$~\GeVc. The mean, $\langle\ktyabs\rangle$, is compared to the earlier presented PYTHIA8 tune in Sec. \ref{sec:EvoFullJetPt} and is in agreement within the uncertainties of the measurement. The mean for $60<\ptjetfull<80$~\GeVc{} is reported for two multiplicity event classes in Tab.~\ref{tab:MomentsV0AEventClass}. No significant difference is observed as a function of the multiplicity measured with V0A.

\section{Conclusion}
The dijet acoplanarity in \pPb{} collisions was studied by measuring dijet transverse momentum \ktyabs. The evolution of \ktyabs{} as function of the transverse momentum of the full jet, associated charged jet and event multiplicity was presented. The \ktyabs{} spectra for different full and associated charged jet transverse momentum ranges in the 0-40\% V0A event multiplicity class were found consistent with the PYTHIA prediction. The observed increase with jet energy from the mean \ktyabs{} of $6.6\pm0.4\rom{(stat.)}\pm0.2\rom{(syst.)}$~\GeVc{} to $18.8\pm1.3\rom{(stat.)}\pm1.5\rom{(syst.)}$~\GeVc{} as well as the observed narrowing of \ktyabs{} for more balanced jets suggests that the dijet \ktyabs{} spectrum for large $Q^2$ processes is mainly sensitive to the increased available phase-space for QCD radiation processes. Furthermore the dijet acoplanarity was found to be consistent (within $1.2\sigma$) in the two event multiplicity classes analyzed in this study, indicating that in the 
measured kinematical region no strong nuclear matter effects in \pPb{} collisions are observed. Since these results indicate that nuclear \kt{} effects are small, the \pt{} imbalance of jet correlations in Pb-Pb results \cite{Aad:2010bu,Chatrchyan:2011sx} are unlikely to originate from multiple scatterings in the nuclear target.

%%%%%%%% acknowledgements
\newenvironment{acknowledgement}{\relax}{\relax}
\begin{acknowledgement}
\section*{Acknowledgements}
% $Id: acknowledgements.tex 1845 2015-01-14 18:51:22Z loizides $
% Version: Jan 2015

The ALICE Collaboration would like to thank all its engineers and technicians for their invaluable contributions to the construction of the experiment and the CERN accelerator teams for the outstanding performance of the LHC complex.
The ALICE Collaboration gratefully acknowledges the resources and support provided by all Grid centres and the Worldwide LHC Computing Grid (WLCG) collaboration.
The ALICE Collaboration acknowledges the following funding agencies for their support in building and
running the ALICE detector:
State Committee of Science,  World Federation of Scientists (WFS)
and Swiss Fonds Kidagan, Armenia,
Conselho Nacional de Desenvolvimento Cient\'{\i}fico e Tecnol\'{o}gico (CNPq), Financiadora de Estudos e Projetos (FINEP),
Funda\c{c}\~{a}o de Amparo \`{a} Pesquisa do Estado de S\~{a}o Paulo (FAPESP);
National Natural Science Foundation of China (NSFC), the Chinese Ministry of Education (CMOE)
and the Ministry of Science and Technology of China (MSTC);
Ministry of Education and Youth of the Czech Republic;
Danish Natural Science Research Council, the Carlsberg Foundation and the Danish National Research Foundation;
The European Research Council under the European Community's Seventh Framework Programme;
Helsinki Institute of Physics and the Academy of Finland;
French CNRS-IN2P3, the `Region Pays de Loire', `Region Alsace', `Region Auvergne' and CEA, France;
German Bundesministerium fur Bildung, Wissenschaft, Forschung und Technologie (BMBF) and the Helmholtz Association;
General Secretariat for Research and Technology, Ministry of
Development, Greece;
Hungarian Orszagos Tudomanyos Kutatasi Alappgrammok (OTKA) and National Office for Research and Technology (NKTH);
Department of Atomic Energy and Department of Science and Technology of the Government of India;
Istituto Nazionale di Fisica Nucleare (INFN) and Centro Fermi -
Museo Storico della Fisica e Centro Studi e Ricerche "Enrico
Fermi", Italy;
MEXT Grant-in-Aid for Specially Promoted Research, Ja\-pan;
Joint Institute for Nuclear Research, Dubna;
National Research Foundation of Korea (NRF);
Consejo Nacional de Cienca y Tecnologia (CONACYT), Direccion General de Asuntos del Personal Academico(DGAPA), M\'{e}xico, :Amerique Latine Formation academique – European Commission(ALFA-EC) and the EPLANET Program
(European Particle Physics Latin American Network)
Stichting voor Fundamenteel Onderzoek der Materie (FOM) and the Nederlandse Organisatie voor Wetenschappelijk Onderzoek (NWO), Netherlands;
Research Council of Norway (NFR);
National Science Centre, Poland;
Ministry of National Education/Institute for Atomic Physics and Consiliul Naţional al Cercetării Ştiinţifice - Executive Agency for Higher Education Research Development and Innovation Funding (CNCS-UEFISCDI) - Romania;
Ministry of Education and Science of Russian Federation, Russian
Academy of Sciences, Russian Federal Agency of Atomic Energy,
Russian Federal Agency for Science and Innovations and The Russian
Foundation for Basic Research;
Ministry of Education of Slovakia;
Department of Science and Technology, South Africa;
Centro de Investigaciones Energeticas, Medioambientales y Tecnologicas (CIEMAT), E-Infrastructure shared between Europe and Latin America (EELA), Ministerio de Econom\'{i}a y Competitividad (MINECO) of Spain, Xunta de Galicia (Conseller\'{\i}a de Educaci\'{o}n),
Centro de Aplicaciones Tecnológicas y Desarrollo Nuclear (CEA\-DEN), Cubaenerg\'{\i}a, Cuba, and IAEA (International Atomic Energy Agency);
Swedish Research Council (VR) and Knut $\&$ Alice Wallenberg
Foundation (KAW);
Ukraine Ministry of Education and Science;
United Kingdom Science and Technology Facilities Council (STFC);
The United States Department of Energy, the United States National
Science Foundation, the State of Texas, and the State of Ohio;
Ministry of Science, Education and Sports of Croatia and  Unity through Knowledge Fund, Croatia.
Council of Scientific and Industrial Research (CSIR), New Delhi, India
    %%%%%%% done by webmaster team
\end{acknowledgement}

\bibliographystyle{utphys}
\bibliography{biblio}

\providecommand{\href}[2]{#2}\begingroup\raggedright\begin{thebibliography}{10}

\bibitem{Feynman:1977yr}
R.~Feynman, R.~Field, and G.~Fox, ``{Correlations Among Particles and Jets
  Produced with Large Transverse Momenta}''
\href{http://dx.doi.org/10.1016/0550-3213(77)90299-1}{{\em Nucl.Phys.}
  {\bfseries B128} (1977) 1--65}.
%%CITATION = NUPHA,B128,1;%%.

\bibitem{PhysRevLett.81.2642}
L.~Apanasevich {\em et~al.}, ``Evidence for parton ${k}_{T}$ effects in high-
  ${p}_{T}$ particle production''
  \href{http://dx.doi.org/10.1103/PhysRevLett.81.2642}{{\em Phys. Rev. Lett.}
  {\bfseries 81} (Sep, 1998) 2642--2645}.

\bibitem{PhysRevD.59.074007}
L.~Apanasevich {\em et~al.}, ``${k}_{T}$ effects in direct-photon production''
  \href{http://dx.doi.org/10.1103/PhysRevD.59.074007}{{\em Phys. Rev. D}
  {\bfseries 59} (Feb, 1999) 074007}.

\bibitem{Abazov:2004hm}
{\bfseries D0 Collaboration}, V.~Abazov {\em et~al.}, ``{Measurement of dijet
  azimuthal decorrelations at central rapidities in $p\bar{p}$ collisions at
  $\sqrt{s} = 1.96$ TeV}''
  \href{http://dx.doi.org/10.1103/PhysRevLett.94.221801}{{\em Phys.Rev.Lett.}
  {\bfseries 94} (2005) 221801},
\href{http://arxiv.org/abs/hep-ex/0409040}{{\ttfamily arXiv:hep-ex/0409040
  [hep-ex]}}.
%%CITATION = HEP-EX/0409040;%%.

\bibitem{Khachatryan:2011zj}
{\bfseries CMS Collaboration}, V.~Khachatryan {\em et~al.}, ``{Dijet Azimuthal
  Decorrelations in $pp$ Collisions at $\sqrt{s} = 7$~TeV}''
  \href{http://dx.doi.org/10.1103/PhysRevLett.106.122003}{{\em Phys.Rev.Lett.}
  {\bfseries 106} (2011) 122003},
\href{http://arxiv.org/abs/1101.5029}{{\ttfamily arXiv:1101.5029 [hep-ex]}}.
%%CITATION = ARXIV:1101.5029;%%.

\bibitem{daCosta:2011ni}
{\bfseries ATLAS Collaboration}, G.~Aad {\em et~al.}, ``{Measurement of Dijet
  Azimuthal Decorrelations in $pp$ Collisions at $\sqrt{s}=7$ TeV}''
  \href{http://dx.doi.org/10.1103/PhysRevLett.106.172002}{{\em Phys.Rev.Lett.}
  {\bfseries 106} (2011) 172002},
\href{http://arxiv.org/abs/1102.2696}{{\ttfamily arXiv:1102.2696 [hep-ex]}}.
%%CITATION = ARXIV:1102.2696;%%.

\bibitem{Albacete:2013ei}
J.~Albacete, N.~Armesto, R.~Baier, G.~Barnafoldi, J.~Barrette, {\em et~al.},
  ``{Predictions for p+Pb Collisions at $\sqrt{s_{NN}}$ = 5 TeV}''
  \href{http://dx.doi.org/10.1142/S0218301313300075}{{\em Int.J.Mod.Phys.}
  {\bfseries E22} (2013) 1330007},
\href{http://arxiv.org/abs/1301.3395}{{\ttfamily arXiv:1301.3395 [hep-ph]}}.
%%CITATION = ARXIV:1301.3395;%%.

\bibitem{Gyulassy:1990ye}
M.~Gyulassy and M.~Plumer, ``{Jet quenching in dense matter}''
\href{http://dx.doi.org/10.1016/0370-2693(90)91409-5}{{\em Phys.Lett.}
  {\bfseries B243} (1990) 432--438}.
%%CITATION = PHLTA,B243,432;%%.

\bibitem{Baier:1994bd}
R.~Baier, Y.~L. Dokshitzer, S.~Peigne, and D.~Schiff, ``{{Induced gluon
  radiation in a QCD medium}}''
  \href{http://dx.doi.org/10.1016/0370-2693(94)01617-L}{{\em Phys.Lett.}
  {\bfseries B345} (1995) 277--286},
\href{http://arxiv.org/abs/hep-ph/9411409}{{\ttfamily arXiv:hep-ph/9411409
  [hep-ph]}}.
%%CITATION = HEP-PH/9411409;%%.

\bibitem{Aad:2014bxa}
{\bfseries ATLAS Collaboration}, G.~Aad {\em et~al.}, ``{Measurements of the
  Nuclear Modification Factor for Jets in Pb+Pb Collisions at
  $\sqrt{s_{\mathrm{NN}}}=2.76$ TeV with the ATLAS Detector}''
\href{http://arxiv.org/abs/1411.2357}{{\ttfamily arXiv:1411.2357 [hep-ex]}}.
%%CITATION = ARXIV:1411.2357;%%.

\bibitem{Aad:2014wha}
{\bfseries ATLAS Collaboration}, G.~Aad {\em et~al.}, ``{Measurement of
  inclusive jet charged-particle fragmentation functions in Pb+Pb collisions at
  $\sqrt{s_{NN}} = 2.76$ TeV with the ATLAS detector}''
  \href{http://dx.doi.org/10.1016/j.physletb.2014.10.065}{{\em Phys.Lett.}
  {\bfseries B739} (2014) 320--342},
\href{http://arxiv.org/abs/1406.2979}{{\ttfamily arXiv:1406.2979 [hep-ex]}}.
%%CITATION = ARXIV:1406.2979;%%.

\bibitem{Chatrchyan:2014ava}
{\bfseries CMS Collaboration}, S.~Chatrchyan {\em et~al.}, ``{Measurement of
  jet fragmentation in PbPb and pp collisions at $\sqrt{s_{NN}}=2.76$ TeV}''
  \href{http://dx.doi.org/10.1103/PhysRevC.90.024908}{{\em Phys.Rev.}
  {\bfseries C90} no.~2, (2014) 024908},
\href{http://arxiv.org/abs/1406.0932}{{\ttfamily arXiv:1406.0932 [nucl-ex]}}.
%%CITATION = ARXIV:1406.0932;%%.

\bibitem{Chatrchyan:2013kwa}
{\bfseries CMS Collaboration}, S.~Chatrchyan {\em et~al.}, ``{Modification of
  jet shapes in PbPb collisions at $\sqrt {s_{NN}} = 2.76$ TeV}''
  \href{http://dx.doi.org/10.1016/j.physletb.2014.01.042}{{\em Phys.Lett.}
  {\bfseries B730} (2014) 243--263},
\href{http://arxiv.org/abs/1310.0878}{{\ttfamily arXiv:1310.0878 [nucl-ex]}}.
%%CITATION = ARXIV:1310.0878;%%.

\bibitem{Chatrchyan:2014hqa}
{\bfseries CMS Collaboration}, S.~Chatrchyan {\em et~al.}, ``{Studies of dijet
  transverse momentum balance and pseudorapidity distributions in pPb
  collisions at $\sqrt{s_{\mathrm{NN}}} = 5.02$ $\,\text {TeV}$}''
  \href{http://dx.doi.org/10.1140/epjc/s10052-014-2951-y}{{\em Eur.Phys.J.}
  {\bfseries C74} no.~7, (2014) 2951},
\href{http://arxiv.org/abs/1401.4433}{{\ttfamily arXiv:1401.4433 [nucl-ex]}}.
%%CITATION = ARXIV:1401.4433;%%.

\bibitem{Adam:2014qja}
{\bfseries ALICE Collaboration}, J.~Adam {\em et~al.}, ``{Centrality dependence
  of particle production in p-Pb collisions at $\sqrt{s_{\rm NN} }$= 5.02
  TeV}''
\href{http://arxiv.org/abs/1412.6828}{{\ttfamily arXiv:1412.6828 [nucl-ex]}}.
%%CITATION = ARXIV:1412.6828;%%.

\bibitem{ALICE:2012xs}
{\bfseries ALICE Collaboration}, B.~Abelev {\em et~al.}, ``{Pseudorapidity
  density of charged particles in $p$-Pb at $\sqrt{s_{NN}}=5.02$ TeV}''
  \href{http://dx.doi.org/10.1103/PhysRevLett.110.032301}{{\em Phys.Rev.Lett.}
  {\bfseries 110} (2013) 032301},
\href{http://arxiv.org/abs/1210.3615}{{\ttfamily arXiv:1210.3615 [nucl-ex]}}.
%%CITATION = ARXIV:1210.3615;%%.

\bibitem{oai:arXiv.org:1008.0413}
{\bfseries ALICE EMCal Collaboration}, U.~Abeysekara {\em et~al.}, ``{ALICE
  EMCal Physics Performance Report}''
\href{http://arxiv.org/abs/1008.0413}{{\ttfamily arXiv:1008.0413
  [physics.ins-det]}}.
%%CITATION = ARXIV:1008.0413;%%.

\bibitem{Aamodt:2010aa}
{\bfseries ALICE Collaboration}, K.~Aamodt {\em et~al.}, ``{Alignment of the
  ALICE Inner Tracking System with cosmic-ray tracks}''
  \href{http://dx.doi.org/10.1088/1748-0221/5/03/P03003}{{\em JINST} {\bfseries
  5} (2010) P03003},
\href{http://arxiv.org/abs/1001.0502}{{\ttfamily arXiv:1001.0502
  [physics.ins-det]}}.
%%CITATION = ARXIV:1001.0502;%%.

\bibitem{Alme:2010ke}
J.~Alme, Y.~Andres, H.~Appelshauser, S.~Bablok, N.~Bialas, {\em et~al.}, ``{The
  ALICE TPC, a large 3-dimensional tracking device with fast readout for
  ultra-high multiplicity events}''
  \href{http://dx.doi.org/10.1016/j.nima.2010.04.042}{{\em Nucl.Instrum.Meth.}
  {\bfseries A622} (2010) 316--367},
\href{http://arxiv.org/abs/1001.1950}{{\ttfamily arXiv:1001.1950
  [physics.ins-det]}}.
%%CITATION = ARXIV:1001.1950;%%.

\bibitem{Abelev:2014ffa}
{\bfseries ALICE Collaboration}, B.~Abelev {\em et~al.}, ``{Performance of the
  ALICE Experiment at the CERN LHC}'' {\em Int.J.Mod.Phys.} {\bfseries A29}
  (2014) 1430044,
\href{http://arxiv.org/abs/1402.4476}{{\ttfamily arXiv:1402.4476 [nucl-ex]}}.
%%CITATION = ARXIV:1402.4476;%%.

\bibitem{Abelev:2014mva}
{\bfseries ALICE Collaboration}, B.~Abelev {\em et~al.}, ``{Multiplicity
  dependence of jet-like two-particle correlations in p-Pb collisions at
  $\sqrt{s_{NN}}$ = 5.02 TeV}''
  \href{http://dx.doi.org/10.1016/j.physletb.2014.11.028}{{\em Phys.Lett.}
  {\bfseries B741} (2014) 38--50},
\href{http://arxiv.org/abs/1406.5463}{{\ttfamily arXiv:1406.5463 [nucl-ex]}}.
%%CITATION = ARXIV:1406.5463;%%.

\bibitem{Cacciari2011}
M.~Cacciari, G.~P. Salam, and G.~Soyez, ``{FastJet User Manual}''
  \href{http://dx.doi.org/10.1140/epjc/s10052-012-1896-2}{{\em Eur.Phys.J.}
  {\bfseries C72} (2012) 1896},
\href{http://arxiv.org/abs/1111.6097}{{\ttfamily arXiv:1111.6097 [hep-ph]}}.
%%CITATION = ARXIV:1111.6097;%%.

\bibitem{Cacciari2006}
M.~Cacciari and G.~P. Salam, ``Dispelling the $n^3$ myth for the
  $k_{\mathrm{t}}$ jet-finder'' {\em Phys.Lett.B} {\bfseries 641} (2006)
  57--61, \href{http://arxiv.org/abs/hep-ph/0512210}{{\ttfamily
  hep-ph/0512210}}.

\bibitem{Abelev:2012ej}
{\bfseries ALICE Collaboration}, B.~Abelev {\em et~al.}, ``{Measurement of
  Event Background Fluctuations for Charged Particle Jet Reconstruction in
  Pb-Pb collisions at $\sqrt{s_{NN}} = 2.76$ TeV}''
  \href{http://dx.doi.org/10.1007/JHEP03(2012)053}{{\em JHEP} {\bfseries 1203}
  (2012) 053},
\href{http://arxiv.org/abs/1201.2423}{{\ttfamily arXiv:1201.2423 [hep-ex]}}.
%%CITATION = ARXIV:1201.2423;%%.

\bibitem{Abelev:2013fn}
{\bfseries ALICE Collaboration}, B.~Abelev {\em et~al.}, ``{Measurement of the
  inclusive differential jet cross section in $pp$ collisions at $\sqrt{s} =
  2.76$ TeV}'' \href{http://dx.doi.org/10.1016/j.physletb.2013.04.026}{{\em
  Phys.Lett.} {\bfseries B722} (2013) 262--272},
\href{http://arxiv.org/abs/1301.3475}{{\ttfamily arXiv:1301.3475 [nucl-ex]}}.
%%CITATION = ARXIV:1301.3475;%%.

\bibitem{Chatrchyan:2012tt}
{\bfseries CMS Collaboration}, S.~Chatrchyan {\em et~al.}, ``{Measurement of
  the underlying event activity in $pp$ collisions at $\sqrt{s} = 0.9$ and 7
  TeV with the novel jet-area/median approach}''
  \href{http://dx.doi.org/10.1007/JHEP08(2012)130}{{\em JHEP} {\bfseries 1208}
  (2012) 130},
\href{http://arxiv.org/abs/1207.2392}{{\ttfamily arXiv:1207.2392 [hep-ex]}}.
%%CITATION = ARXIV:1207.2392;%%.

\bibitem{Adam:2015hoa}
{\bfseries ALICE Collaboration}, J.~Adam {\em et~al.}, ``{Measurement of
  charged jet production cross sections and nuclear modification in p-Pb
  collisions at $\sqrt{s_\mathrm{NN}} = 5.02$ TeV}''
\href{http://arxiv.org/abs/1503.00681}{{\ttfamily arXiv:1503.00681 [nucl-ex]}}.
%%CITATION = ARXIV:1503.00681;%%.

\bibitem{Angelis1980163}
A.~Angelis~et al., ``A measurement of the transverse momenta of partons, and of
  jet fragmentation as a function of $\sqrt{s}$ in p-p collisions''
  \href{http://dx.doi.org/http://dx.doi.org/10.1016/0370-2693(80)90572-9}{{\em
  Physics Letters B} {\bfseries 97} no.~1, (1980) 163 -- 168}.

\bibitem{Sjostrand2006}
T.~Sjostrand, S.~Mrenna, and P.~Skands, ``{PYTHIA} 6.4 physics and manual''
  {\em JHEP} {\bfseries 05} (2006) 026,
  \href{http://arxiv.org/abs/hep-ph/0603175}{{\ttfamily hep-ph/0603175}}.

\bibitem{Sjostrand:2007gs}
T.~Sjostrand, S.~Mrenna, and P.~Z. Skands, ``{A Brief Introduction to PYTHIA
  8.1}'' \href{http://dx.doi.org/10.1016/j.cpc.2008.01.036}{{\em
  Comput.Phys.Commun.} {\bfseries 178} (2008) 852--867},
\href{http://arxiv.org/abs/0710.3820}{{\ttfamily arXiv:0710.3820 [hep-ph]}}.
%%CITATION = ARXIV:0710.3820;%%.

\bibitem{Skands:2010ak}
P.~Z. Skands, ``{Tuning Monte Carlo Generators: The Perugia Tunes}''
  \href{http://dx.doi.org/10.1103/PhysRevD.82.074018}{{\em Phys.Rev.}
  {\bfseries D82} (2010) 074018},
\href{http://arxiv.org/abs/1005.3457}{{\ttfamily arXiv:1005.3457 [hep-ph]}}.
%%CITATION = ARXIV:1005.3457;%%.

\bibitem{Brun1994}
R.~Brun, F.~Carminati, and S.~Giani,
``{GEANT Detector Description and Simulation Tool}''.
%%CITATION = CERN-W5013 ETC.;%%.

\bibitem{Abelev:2013kqa}
{\bfseries ALICE Collaboration}, B.~Abelev {\em et~al.}, ``{Measurement of
  charged jet suppression in Pb-Pb collisions at $\sqrt{s_{NN}}$ = 2.76 TeV}''
  \href{http://dx.doi.org/10.1007/JHEP03(2014)013}{{\em JHEP} {\bfseries 1403}
  (2014) 013},
\href{http://arxiv.org/abs/1311.0633}{{\ttfamily arXiv:1311.0633 [nucl-ex]}}.
%%CITATION = ARXIV:1311.0633;%%.

\bibitem{ALICE:2014dla}
{\bfseries ALICE Collaboration}, B.~B. Abelev {\em et~al.}, ``{Charged jet
  cross sections and properties in proton-proton collisions at $\sqrt{s}$ = 7
  TeV}''
\href{http://arxiv.org/abs/1411.4969}{{\ttfamily arXiv:1411.4969 [nucl-ex]}}.
%%CITATION = ARXIV:1411.4969;%%.

\bibitem{Aad:2010bu}
{\bfseries ATLAS Collaboration}, G.~Aad {\em et~al.}, ``{Observation of a
  Centrality-Dependent Dijet Asymmetry in Lead-Lead Collisions at
  $\sqrt{s_{NN}}=2.77$ TeV with the ATLAS Detector at the LHC}''
  \href{http://dx.doi.org/10.1103/PhysRevLett.105.252303}{{\em Phys.Rev.Lett.}
  {\bfseries 105} (2010) 252303},
\href{http://arxiv.org/abs/1011.6182}{{\ttfamily arXiv:1011.6182 [hep-ex]}}.
%%CITATION = ARXIV:1011.6182;%%.

\bibitem{Chatrchyan:2011sx}
{\bfseries CMS Collaboration}, S.~Chatrchyan {\em et~al.}, ``{Observation and
  studies of jet quenching in PbPb collisions at nucleon-nucleon center-of-mass
  energy = 2.76 TeV}'' \href{http://dx.doi.org/10.1103/PhysRevC.84.024906}{{\em
  Phys.Rev.} {\bfseries C84} (2011) 024906},
\href{http://arxiv.org/abs/1102.1957}{{\ttfamily arXiv:1102.1957 [nucl-ex]}}.
%%CITATION = ARXIV:1102.1957;%%.

\end{thebibliography}\endgroup

%%%%%%%%% appendix with author list
\newpage
\appendix
\section{The ALICE Collaboration}\label{app:collab}

% Collaboration: CERN-LHC-ALICE
% Generation Date is 2015/Feb/18

% How to use:
%%%%%%%%% appendix with author list
%\appendix
%\section{The ALICE Collaboration}
%\label{app:collab}
%\input{authors-list.tex}  %%%%%%% get the latest version before submitting

\begingroup
\small
\begin{flushleft}
J.~Adam\Irefn{org39}\And
D.~Adamov\'{a}\Irefn{org82}\And
M.M.~Aggarwal\Irefn{org86}\And
G.~Aglieri Rinella\Irefn{org36}\And
M.~Agnello\Irefn{org110}\And
N.~Agrawal\Irefn{org47}\And
Z.~Ahammed\Irefn{org130}\And
S.U.~Ahn\Irefn{org67}\And
I.~Aimo\Irefn{org93}\textsuperscript{,}\Irefn{org110}\And
S.~Aiola\Irefn{org135}
% ...
\renewcommand\labelenumi{\textsuperscript{\theenumi}~}

\section*{Affiliation notes}
\renewcommand\theenumi{\roman{enumi}}
\begin{Authlist}
\item \Adef{0}Deceased
\item \Adef{idp5843168}{Also at: University of Kansas, Lawrence, Kansas, United States}
\end{Authlist}

\section*{Collaboration Institutes}
\renewcommand\theenumi{\arabic{enumi}~}
\begin{Authlist}

\item \Idef{org1}A.I. Alikhanyan National Science Laboratory (Yerevan Physics Institute) Foundation, Yerevan, Armenia
\item \Idef{org2}Benem\'{e}rita Universidad Aut\'{o}noma de Puebla, Puebla, Mexico
\item \Idef{org3}Bogolyubov Institute for Theoretical Physics, Kiev, Ukraine
\item \Idef{org4}Bose Institute, Department of Physics and Centre for Astroparticle Physics and Space Science (CAPSS), Kolkata, India
\item \Idef{org5}Budker Institute for Nuclear Physics, Novosibirsk, Russia
\item \Idef{org6}California Polytechnic State University, San Luis Obispo, California, United States
\item \Idef{org7}Central China Normal University, Wuhan, China
\item \Idef{org8}Centre de Calcul de l'IN2P3, Villeurbanne, France
\item \Idef{org9}Centro de Aplicaciones Tecnol\'{o}gicas y Desarrollo Nuclear (CEADEN), Havana, Cuba
\item \Idef{org10}Centro de Investigaciones Energ\'{e}ticas Medioambientales y Tecnol\'{o}gicas (CIEMAT), Madrid, Spain
\item \Idef{org11}Centro de Investigaci\'{o}n y de Estudios Avanzados (CINVESTAV), Mexico City and M\'{e}rida, Mexico
\item \Idef{org12}Centro Fermi - Museo Storico della Fisica e Centro Studi e Ricerche ``Enrico Fermi'', Rome, Italy
\item \Idef{org13}Chicago State University, Chicago, Illinois, USA
\item \Idef{org14}China Institute of Atomic Energy, Beijing, China
\item \Idef{org15}Commissariat \`{a} l'Energie Atomique, IRFU, Saclay, France
\item \Idef{org16}COMSATS Institute of Information Technology (CIIT), Islamabad, Pakistan
\item \Idef{org17}Departamento de F\'{\i}sica de Part\'{\i}culas and IGFAE, Universidad de Santiago de Compostela, Santiago de Compostela, Spain
\item \Idef{org18}Department of Physics and Technology, University of Bergen, Bergen, Norway
\item \Idef{org19}Department of Physics, Aligarh Muslim University, Aligarh, India
\item \Idef{org20}Department of Physics, Ohio State University, Columbus, Ohio, United States
\item \Idef{org21}Department of Physics, Sejong University, Seoul, South Korea
\item \Idef{org22}Department of Physics, University of Oslo, Oslo, Norway
\item \Idef{org23}Dipartimento di Elettrotecnica ed Elettronica del Politecnico, Bari, Italy
\item \Idef{org24}Dipartimento di Fisica dell'Universit\`{a} 'La Sapienza' and Sezione INFN Rome, Italy
\item \Idef{org25}Dipartimento di Fisica dell'Universit\`{a} and Sezione INFN, Cagliari, Italy
\item \Idef{org26}Dipartimento di Fisica dell'Universit\`{a} and Sezione INFN, Trieste, Italy
\item \Idef{org27}Dipartimento di Fisica dell'Universit\`{a} and Sezione INFN, Turin, Italy
\item \Idef{org28}Dipartimento di Fisica e Astronomia dell'Universit\`{a} and Sezione INFN, Bologna, Italy
\item \Idef{org29}Dipartimento di Fisica e Astronomia dell'Universit\`{a} and Sezione INFN, Catania, Italy
\item \Idef{org30}Dipartimento di Fisica e Astronomia dell'Universit\`{a} and Sezione INFN, Padova, Italy
\item \Idef{org31}Dipartimento di Fisica `E.R.~Caianiello' dell'Universit\`{a} and Gruppo Collegato INFN, Salerno, Italy
\item \Idef{org32}Dipartimento di Scienze e Innovazione Tecnologica dell'Universit\`{a} del  Piemonte Orientale and Gruppo Collegato INFN, Alessandria, Italy
\item \Idef{org33}Dipartimento Interateneo di Fisica `M.~Merlin' and Sezione INFN, Bari, Italy
\item \Idef{org34}Division of Experimental High Energy Physics, University of Lund, Lund, Sweden
\item \Idef{org35}Eberhard Karls Universit\"{a}t T\"{u}bingen, T\"{u}bingen, Germany
\item \Idef{org36}European Organization for Nuclear Research (CERN), Geneva, Switzerland
\item \Idef{org37}Faculty of Engineering, Bergen University College, Bergen, Norway
\item \Idef{org38}Faculty of Mathematics, Physics and Informatics, Comenius University, Bratislava, Slovakia
\item \Idef{org39}Faculty of Nuclear Sciences and Physical Engineering, Czech Technical University in Prague, Prague, Czech Republic
\item \Idef{org40}Faculty of Science, P.J.~\v{S}af\'{a}rik University, Ko\v{s}ice, Slovakia
\item \Idef{org41}Faculty of Technology, Buskerud and Vestfold University College, Vestfold, Norway
\item \Idef{org42}Frankfurt Institute for Advanced Studies, Johann Wolfgang Goethe-Universit\"{a}t Frankfurt, Frankfurt, Germany
\item \Idef{org43}Gangneung-Wonju National University, Gangneung, South Korea
\item \Idef{org44}Gauhati University, Department of Physics, Guwahati, India
\item \Idef{org45}Helsinki Institute of Physics (HIP), Helsinki, Finland
\item \Idef{org46}Hiroshima University, Hiroshima, Japan
\item \Idef{org47}Indian Institute of Technology Bombay (IIT), Mumbai, India
\item \Idef{org48}Indian Institute of Technology Indore, Indore (IITI), India
\item \Idef{org49}Inha University, Incheon, South Korea
\item \Idef{org50}Institut de Physique Nucl\'eaire d'Orsay (IPNO), Universit\'e Paris-Sud, CNRS-IN2P3, Orsay, France
\item \Idef{org51}Institut f\"{u}r Informatik, Johann Wolfgang Goethe-Universit\"{a}t Frankfurt, Frankfurt, Germany
\item \Idef{org52}Institut f\"{u}r Kernphysik, Johann Wolfgang Goethe-Universit\"{a}t Frankfurt, Frankfurt, Germany
\item \Idef{org53}Institut f\"{u}r Kernphysik, Westf\"{a}lische Wilhelms-Universit\"{a}t M\"{u}nster, M\"{u}nster, Germany
\item \Idef{org54}Institut Pluridisciplinaire Hubert Curien (IPHC), Universit\'{e} de Strasbourg, CNRS-IN2P3, Strasbourg, France
\item \Idef{org55}Institute for Nuclear Research, Academy of Sciences, Moscow, Russia
\item \Idef{org56}Institute for Subatomic Physics of Utrecht University, Utrecht, Netherlands
\item \Idef{org57}Institute for Theoretical and Experimental Physics, Moscow, Russia
\item \Idef{org58}Institute of Experimental Physics, Slovak Academy of Sciences, Ko\v{s}ice, Slovakia
\item \Idef{org59}Institute of Physics, Academy of Sciences of the Czech Republic, Prague, Czech Republic
\item \Idef{org60}Institute of Physics, Bhubaneswar, India
\item \Idef{org61}Institute of Space Science (ISS), Bucharest, Romania
\item \Idef{org62}Instituto de Ciencias Nucleares, Universidad Nacional Aut\'{o}noma de M\'{e}xico, Mexico City, Mexico
\item \Idef{org63}Instituto de F\'{\i}sica, Universidad Nacional Aut\'{o}noma de M\'{e}xico, Mexico City, Mexico
\item \Idef{org64}iThemba LABS, National Research Foundation, Somerset West, South Africa
\item \Idef{org65}Joint Institute for Nuclear Research (JINR), Dubna, Russia
\item \Idef{org66}Konkuk University, Seoul, South Korea
\item \Idef{org67}Korea Institute of Science and Technology Information, Daejeon, South Korea
\item \Idef{org68}KTO Karatay University, Konya, Turkey
\item \Idef{org69}Laboratoire de Physique Corpusculaire (LPC), Clermont Universit\'{e}, Universit\'{e} Blaise Pascal, CNRS--IN2P3, Clermont-Ferrand, France
\item \Idef{org70}Laboratoire de Physique Subatomique et de Cosmologie, Universit\'{e} Grenoble-Alpes, CNRS-IN2P3, Grenoble, France
\item \Idef{org71}Laboratori Nazionali di Frascati, INFN, Frascati, Italy
\item \Idef{org72}Laboratori Nazionali di Legnaro, INFN, Legnaro, Italy
\item \Idef{org73}Lawrence Berkeley National Laboratory, Berkeley, California, United States
\item \Idef{org74}Lawrence Livermore National Laboratory, Livermore, California, United States
\item \Idef{org75}Moscow Engineering Physics Institute, Moscow, Russia
\item \Idef{org76}National Centre for Nuclear Studies, Warsaw, Poland
\item \Idef{org77}National Institute for Physics and Nuclear Engineering, Bucharest, Romania
\item \Idef{org78}National Institute of Science Education and Research, Bhubaneswar, India
\item \Idef{org79}Niels Bohr Institute, University of Copenhagen, Copenhagen, Denmark
\item \Idef{org80}Nikhef, National Institute for Subatomic Physics, Amsterdam, Netherlands
\item \Idef{org81}Nuclear Physics Group, STFC Daresbury Laboratory, Daresbury, United Kingdom
\item \Idef{org82}Nuclear Physics Institute, Academy of Sciences of the Czech Republic, \v{R}e\v{z} u Prahy, Czech Republic
\item \Idef{org83}Oak Ridge National Laboratory, Oak Ridge, Tennessee, United States
\item \Idef{org84}Petersburg Nuclear Physics Institute, Gatchina, Russia
\item \Idef{org85}Physics Department, Creighton University, Omaha, Nebraska, United States
\item \Idef{org86}Physics Department, Panjab University, Chandigarh, India
\item \Idef{org87}Physics Department, University of Athens, Athens, Greece
\item \Idef{org88}Physics Department, University of Cape Town, Cape Town, South Africa
\item \Idef{org89}Physics Department, University of Jammu, Jammu, India
\item \Idef{org90}Physics Department, University of Rajasthan, Jaipur, India
\item \Idef{org91}Physik Department, Technische Universit\"{a}t M\"{u}nchen, Munich, Germany
\item \Idef{org92}Physikalisches Institut, Ruprecht-Karls-Universit\"{a}t Heidelberg, Heidelberg, Germany
\item \Idef{org93}Politecnico di Torino, Turin, Italy
\item \Idef{org94}Purdue University, West Lafayette, Indiana, United States
\item \Idef{org95}Pusan National University, Pusan, South Korea
\item \Idef{org96}Research Division and ExtreMe Matter Institute EMMI, GSI Helmholtzzentrum f\"ur Schwerionenforschung, Darmstadt, Germany
\item \Idef{org97}Rudjer Bo\v{s}kovi\'{c} Institute, Zagreb, Croatia
\item \Idef{org98}Russian Federal Nuclear Center (VNIIEF), Sarov, Russia
\item \Idef{org99}Russian Research Centre Kurchatov Institute, Moscow, Russia
\item \Idef{org100}Saha Institute of Nuclear Physics, Kolkata, India
\item \Idef{org101}School of Physics and Astronomy, University of Birmingham, Birmingham, United Kingdom
\item \Idef{org102}Secci\'{o}n F\'{\i}sica, Departamento de Ciencias, Pontificia Universidad Cat\'{o}lica del Per\'{u}, Lima, Peru
\item \Idef{org103}Sezione INFN, Bari, Italy
\item \Idef{org104}Sezione INFN, Bologna, Italy
\item \Idef{org105}Sezione INFN, Cagliari, Italy
\item \Idef{org106}Sezione INFN, Catania, Italy
\item \Idef{org107}Sezione INFN, Padova, Italy
\item \Idef{org108}Sezione INFN, Rome, Italy
\item \Idef{org109}Sezione INFN, Trieste, Italy
\item \Idef{org110}Sezione INFN, Turin, Italy
\item \Idef{org111}SSC IHEP of NRC Kurchatov institute, Protvino, Russia
\item \Idef{org112}SUBATECH, Ecole des Mines de Nantes, Universit\'{e} de Nantes, CNRS-IN2P3, Nantes, France
\item \Idef{org113}Suranaree University of Technology, Nakhon Ratchasima, Thailand
\item \Idef{org114}Technical University of Split FESB, Split, Croatia
\item \Idef{org115}The Henryk Niewodniczanski Institute of Nuclear Physics, Polish Academy of Sciences, Cracow, Poland
\item \Idef{org116}The University of Texas at Austin, Physics Department, Austin, Texas, USA
\item \Idef{org117}Universidad Aut\'{o}noma de Sinaloa, Culiac\'{a}n, Mexico
\item \Idef{org118}Universidade de S\~{a}o Paulo (USP), S\~{a}o Paulo, Brazil
\item \Idef{org119}Universidade Estadual de Campinas (UNICAMP), Campinas, Brazil
\item \Idef{org120}University of Houston, Houston, Texas, United States
\item \Idef{org121}University of Jyv\"{a}skyl\"{a}, Jyv\"{a}skyl\"{a}, Finland
\item \Idef{org122}University of Liverpool, Liverpool, United Kingdom
\item \Idef{org123}University of Tennessee, Knoxville, Tennessee, United States
\item \Idef{org124}University of the Witwatersrand, Johannesburg, South Africa
\item \Idef{org125}University of Tokyo, Tokyo, Japan
\item \Idef{org126}University of Tsukuba, Tsukuba, Japan
\item \Idef{org127}University of Zagreb, Zagreb, Croatia
\item \Idef{org128}Universit\'{e} de Lyon, Universit\'{e} Lyon 1, CNRS/IN2P3, IPN-Lyon, Villeurbanne, France
\item \Idef{org129}V.~Fock Institute for Physics, St. Petersburg State University, St. Petersburg, Russia
\item \Idef{org130}Variable Energy Cyclotron Centre, Kolkata, India
\item \Idef{org131}Vin\v{c}a Institute of Nuclear Sciences, Belgrade, Serbia
\item \Idef{org132}Warsaw University of Technology, Warsaw, Poland
\item \Idef{org133}Wayne State University, Detroit, Michigan, United States
\item \Idef{org134}Wigner Research Centre for Physics, Hungarian Academy of Sciences, Budapest, Hungary
\item \Idef{org135}Yale University, New Haven, Connecticut, United States
\item \Idef{org136}Yonsei University, Seoul, South Korea
\item \Idef{org137}Zentrum f\"{u}r Technologietransfer und Telekommunikation (ZTT), Fachhochschule Worms, Worms, Germany
\end{Authlist}
\endgroup

  %%%%%%% get the latest version before submitting

\end{document}